\documentclass[preprint2]{aastex}

\begin{document}

%% LaTeX will automatically break titles if they run longer than
%% one line. However, you may use \\ to force a line break if
%% you desire.

\title{Inverse problem in Ionospheric Science: Prediction of solar soft-X-ray spectrum
from Very Low Frequency Radiosonde results}

%% Use \author, \affil, and the \and command to format
%% author and affiliation information.
%% Note that \email has replaced the old \authoremail command
%% from AASTeX v4.0. You can use \email to mark an email address
%% anywhere in the paper, not just in the front matter.
%% As in the title, use \\ to force line breaks.

\author{S. Palit\altaffilmark{1}, S. Ray\altaffilmark{1} and S. K. Chakrabarti\altaffilmark{2,1}}
%%\email{}
%%\affil{Astronomy Department, University of California,
%%    Berkeley, CA 94720}

%%\author{R. Suman\altaffilmark{1}}
%%\email{}

%%\and

%%\author{S. K. Chakrabarti\altaffilmark{1,2}}
%%\email{}

%% Notice that each of these authors has alternate affiliations, which
%% are identified by the \altaffilmark after each name.  Specify alternate
%% affiliation information with \altaffiltext, with one command per each
%% affiliation.

\altaffiltext{1}{Indian Centre for Space Physics, 43-Chalantika, Garia Station Road, Kolkata-700084, India}
\altaffiltext{2}{S. N. Bose National Centre for Basic Sciences, JD Block, Salt Lake, Kolkata-700098, India}

%% Mark off your abstract in the ``abstract'' environment. In the manuscript
%% style, abstract will output a Received/Accepted line after the
%% title and affiliation information. No date will appear since the author
%% does not have this information. The dates will be filled in by the
%% editorial office after submission.

%%\begin{linenumbers}

\begin{abstract}
X-rays and gamma-rays from astronomical sources such as solar flares are mostly absorbed 
by the Earth's atmosphere. Resulting electron-ion production rate as a function of height 
depends on the intensity and wavelength of the injected spectrum and therefore the effects 
vary from one source to another. In other words, the ion density vs. altitude profile 
has the imprint of the incident photon spectrum. In this paper, we investigate whether 
we can invert the problem uniquely by deconvolution of the
VLF amplitude signal to obtain the details of the injected spectrum. We find that it is possible to do 
this up to a certain accuracy. Our method is useful to carry out a similar exercise 
to infer the spectra of more energetic events such as the Gamma Ray Bursts (GRBs), Soft Gamma Ray Repeaters (SGRs)
etc. by probing even the lower part of the atmosphere. We thus show that to certain extent, the Earth's 
atmosphere could be used as a gigantic detector of relatively strong events.

\end{abstract}

%% Keywords should appear after the \end{abstract} command. The uncommented
%% example has been keyed in ApJ style. See the instructions to authors
%% for the journal to which you are submitting your paper to determine
%% what keyword punctuation is appropriate.


%% From the front matter, we move on to the body of the paper.
%% In the first two sections, notice the use of the natbib \citep
%% and \citet commands to identify citations.  The citations are
%% tied to the reference list via symbolic KEYs. The KEY corresponds
%% to the KEY in the \bibitem in the reference list below. We have
%% chosen the first three characters of the first author's name plus
%% the last two numeral of the year of publication as our KEY for
%% each reference.

%% Authors who wish to have the most important objects in their paper
%% linked in the electronic edition to a data center may do so by tagging
%% their objects with \objectname{} or \object{}.  Each macro takes the
%% object name as its required argument. The optional, square-bracket 
%% argument should be used in cases where the data center identification
%% differs from what is to be printed in the paper.  The text appearing 
%% in curly braces is what will appear in print in the published paper. 
%% If the object name is recognized by the data centers, it will be linked
%% in the electronic edition to the object data available at the data centers  
%%
%% Note that for sources with brackets in their names, e.g. [WEG2004] 14h-090,
%% the brackets must be escaped with backslashes when used in the first
%% square-bracket argument, for instance, \object[\[WEG2004\] 14h-090]{90}).
%%  Otherwise, LaTeX will issue an error. 

\section{Introduction}

The study of the effects of solar flares (Whitten \& Poppoff, 1965; Mitra et al., 1972; Grubor et al.,
2005; Xion et al., 2005; Palit et al., 2013) and other
extra-terrestrial high energy phenomena such as Gamma Ray Bursts (GRBs) and Soft Gamma Ray Repeaters (SGRs) 
(Inan et al., 2007; Tanaka et al., 2008; Mondal et al., 2012) on Earth's ionosphere 
is a subject of extended research for last few decades. There have been numerous measurements 
with Very Low Frequency (VLF) Radio waves (Mcrae, 2004; Tanaka et al., 2010; Chakrabarti et al., 2011), 
GPS (Afraimovich et al., 2001; Liu et al., 2004 etc.), RADAR (Taylor \& Watkiks, 1970; 
Watanabe \& Nishitani, 2013) etc. of such effects on different layers of the ionosphere, 
namely, D, E and F regions. These effects, which fall under a more generalized category
called Sudden Ionospheric Disturbances (SIDS) consist mainly of sharp rise and slow decay (flares, GRBs) and
their repetitions (e.g., SGRs). Some models are present in the literature to analyze the effects of ionization 
on the ion-chemistry of the D-region (Mitra \& Rowe, 1972; Turunen et al., 1996; Glukhov et al., 1992) 
and E, F regions (Solomon, 2005; Sojka et al., 2013 etc.). Based on these schemes, successful 
modeling of the effects of solar flares (Palit et al., 2013) and other extra-terrestrial phenomena 
(see, Inan et al., 2007 for $\gamma $ ray flare from Magnetar) on the 
ionosphere has been carried out. With an advancement of knowledge of interaction of solar 
photon with the ionosphere and the chemical processes undergoing there we are now in a position 
to find spectral information of the ionizing events using the upper atmosphere which is a natural
and everlasting detector of high energy photons through its unique response characteristics.  
X-rays and gamma rays of the injected spectra are of great interest mainly for lower ionosphere
or the so-called D-region and regions below it. Lower latitude D-region is affected by X-rays only,
as energetic charged particles from the Sun from the flare events are diverted by Earth's magnetic 
field towards the polar regions. Hence this low and mid latitude ionosphere can be safely used as a `detector'
of high energy photons from extra terrestrial sources. From the ionization  interaction of these photons we can compute the 
response function of this `detector'. Of course, there is no collimation and pointing is very weak 
in the sense that the effects are the highest at the sub-event location and die out away from it. So 
unless there is a single energetic event which obviously dominates the sky, the effects we normally see 
are due to collective events and would be difficult to separate, as in any other detector.

Altitude variation of Electron-ion production rates (Chapman, 1931; Rees, 1989) 
should naturally contain the information of energy distribution of the incoming photons. 
Processes such as recombinations, attachments and detachments etc. with or from ions
and molecules, help electron-ion densities to attain equilibrium values. 
Thus, for obvious reasons, the energy distribution of ionizing X-ray photons during solar flares must 
put its imprints in the form of electron-ion density in the lower ionosphere 
during the occurrence of an event. VLF is found to be only suitable means by which 
the electron density at D-region ionosphere can be estimated on a regular basis. 
Higher frequency radio signals from devices, such as incoherent scatter radars 
are usually very small and prone to be masked by interference and noise. Satellite measurements are unlikely 
as air density at such heights produces too large drag for satellite to float. The height is also beyond the reach of scientific balloons. Thus,
we rely on VLF measurements for electron density distribution measurements.

Variation of electron-ion density affects VLF signal amplitude as it is 
reflected from the D-region. Thus from the change in VLF amplitude, it 
should be possible to extract information about the injected spectra using a
suitable deconvolution processes. In this paper, we have examined this possibility and 
start from the very basic continuity equation inside the D-region or the lower ionosphere. 
In what follows, we present a few such results of deconvolution
and the results of our attempts to find solar flare spectrum from observed VLF modulation data. 
Since VLF is reflected from lower ionosphere or the D-region (in the day time),
we can acquire information about soft X-ray component only during flare events.
Results for other extra-terrestrial sources and for higher energy component of the spectrum will be
presented elsewhere. In a way, our present paper is exactly opposite of what was presented in 
Palit et al. (2013) where we reproduced the observed VLF signals starting from the injected solar spectrum. 

In the next Section, we present the basic theory of how the spectrum 
produces the electron density-height distribution or a function of
it and demonstrate the process of extracting the spectrum information from the distribution profile. 
Section 3 contains our VLF data corresponding to some solar flares 
and how we reproduce the electron density information from the data. In Section 4, 
we present the resulting spectra and compared them with those obtained 
from RHESSI satellite data. In Section 5 we investigate about the accuracy of the ionosphere-detector in providing spectrum information. 
Finally, we discuss limitation of our method and make concluding remarks in the last Section.                         

\section{Basic theory}

\subsection{Continuity equation and the requisite transform}

The dynamics of electron density in the lower ionosphere (D-region in day time) is governed by a simple continuity equation,
$$
\frac {dN_e}{dt} = \frac{q}{1 + \lambda} - \alpha N_e^2,
\eqno{(1)}
$$
where, $N_e$ is the instantaneous electron density, $q$ is the electron-ion production rate by photons which
consists of both the primary photoionizations and ionizations by the photoelectrons. 
$\lambda$ is the ratio of negative ion density and free electron density and
$\alpha$ is the effective recombination coefficient. Both $\lambda$ and $\alpha$ vary 
with time and height and effective variation of $\alpha$ can be calculated accurately (Palit et al. 2013). 
In a normal situation, $q$ mostly consists of contribution from UV ionization. 
During a flare, the ionization in the D-region by X-ray (mainly $\sim 2 - 12$ keV) is much more dominant 
over that due to enhanced UV. Thus, the height variation of $q$ at any instant of time can be assumed to be governed 
by the incident X-ray spectrum only. A simplified form of $q$ due to X-ray as a function 
of time and height can be expressed after a suitable modification of Chapman's formula,
$$
q(h,t) = \sum_j\int I_{0} (\nu, t) e^{{-\sum_k}\sigma_k (\nu)\int_h^{\infty} n_k C_h(h,\psi)dh}
$$
$$
\times \eta_j(\nu)\sigma_j(\nu)n_j(h)d\nu,
\eqno{(2)}
$$
where $I_{0}(\nu,t)d\nu$ is the irradiance or the solar flux at the top of the atmosphere 
in the frequency range $\nu$ to $\nu + d\nu$,
$\sigma_j(\nu)$ is the absorption cross-section for the $j^{th}$ neutral component of air which is a function of energies
of photons. $n_j(h)$ is the concentration and $\eta_j(\nu)$ is the photoionization efficiency for the $j^{th}$
component. $C_h(h,\psi)$ is the grazing incidence function and is given by (Rees, 1989)
$$
\int_h^{\infty} n_j C_h(h,\psi)dh  = \int_h^{\infty} n_j [1 - (\frac {R+h_0}{R+h})^2 sin^2(\psi)]^{\frac {1}{2}}dh
\eqno{(3)}
$$
for $\psi < 90^\circ$.  From Eqs. 1 and Eqs. 2 we get
$$
q(h,t) = \int I_{0}(\nu, t)f(h, \nu)d\nu = (1 + \lambda)(\frac {dN_e}{dt} + \alpha N_e^2),
\eqno{(4)}
$$
where,
$$
f(h, \nu) = \sum_j e^{{-\sum_k}\sigma_k (\nu)\int_h^{\infty} n_k C_h(h,\psi)dh} \eta_j(\nu)\sigma_j(\nu)n_j(\nu)d\nu.
\eqno{(5)}
$$
Clearly Eqs. 4 corresponds to integral transformation between two mathematical functions or spaces ($\nu$ and $h$ spaces) with 
the kernel $f(h,\nu)$ given by Eqs. 5.

\subsection{The basis functions}

The basis function $f(h, \nu)$ in Eqs. 4 is clearly the height profile of electron density produced by a single photon in the energy
range $\nu$ to $\nu + d\nu$. We can find numerically the functions from the knowledge of the X-ray absorption 
cross-sections for different elements from NIST X-ray mass attenuation coefficients table 
(Hubbell and Seltzer, 1995;  www.nist.gov/pml/data/xraycoef) and the concentration 
of neutral elements (MSIS 90 model, Hedin et al., 1991) using the values of $\psi$ 
during the occurrence of the flare. An average value of $\eta_j$ is taken to be $31.4$ 
as obtained in Palit et al. (2013).

\begin{figure}[h]
\centering{
\includegraphics[height=3.5in,width=2.5in,angle = 270]{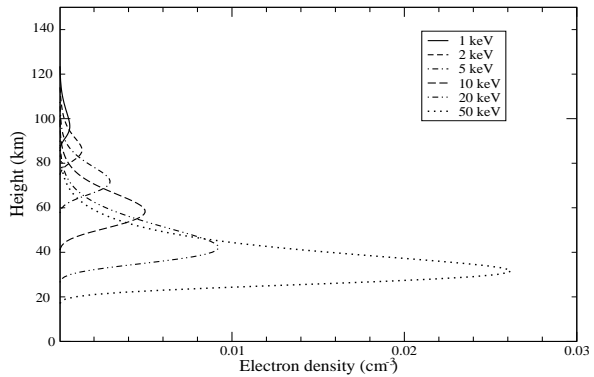}
\caption{A few calculated basis functions $f(h,\nu)$, which correspond
to the electron density produced at a height $h$ due to a 
single incident photon of energy $\frac {h\nu}{c}$(keV).}
\label{fig:basis}}
\end{figure}

Some of the basis vectors are shown in Fig. 1. We find that the contribution 
in electron production in our range of interest ($\sim 60 - 80$ km ) comes 
mainly from photons having energy from $2 - 12$ keV. 
We are interested here in the heights corresponding to the D-region of the 
ionosphere, from where VLF waves are reflected. At this region the values 
of the parameters such as $\lambda$ and $\alpha$ are determined from the interaction of neutral molecules, free electrons 
and three species of ions, namely `Positive', `Negative' and `Positive cluster' ions (Glukhov et al., 1992, Palit et al., 2013).
The photons in higher energy ranges can also contribute in the electron-ion production rate at these heights, only if the abundance 
of photon in those ranges are sufficiently large, but for moderate solar flares this contribution above $\sim 20 keV$ is 
negligible with respect to that from below it. 

For GRBs and SGRs, the contribution from higher energy X-ray and gamma-ray photons should be 
significant and extended to much lower heights (down to $\sim 20 km $). For such 
events we have to take into account the basis corresponding to higher energy photons 
and extend the analysis below the D-region, where the chemistry of interactions is different.
For example, at lower heights ($\sim$ 50 km) negative cluster ions (Inan et al., 2007) 
has to be incorporated in the calculations.

\subsection{Deconvolution to recover desired spectrum}

We present here two possible deconvolution methods, first of which is based on an iterative maximum likelihood estimation
process and the other is based on radial basis function approximations and matrix inversion.

\subsubsection{Iterative Maximum likelihood (modified Richardson Lucy) method }

If for a particular $\nu$, $f(h, \nu)$ is considered as the Point Spread Function (PSF) 
in a space expanded by the values of height, then the spectrum
can be deconvolved from the profile $q(N,\frac{dN}{dt})$ (see Eqs. 4) at any time 
with a Bayesian based iterative Maximum likelihood
estimation method. We choose the Richardson Lucy algorithm 
(Richardson, 1972; Lucy, 1974) and modified it somewhat to suit our case.

The Richardson Lucy algorithm in its most basic form is given by,
$$
s_j^{t+1} = s_j^{t}\sum_i\frac{q_i}{c_i}f_{ij},
\eqno{(6)}
$$
where,
$$
c_i = \sum_jf_{ij}s_j^{t}.
\eqno{(7)}
$$
Here, $f_{ij}$ is the ion density produced in $i^{th}$ height due to a single 
photon from flare in $j^{th}$ keV bin. $q_i$ corresponds to observed density 
distribution at different heights, as calculated from Eqs. 4, $s_j$ is the expected
latent distribution which is to be improved by each iteration ($t$) from a trivially chosen distribution, say $s_j^0$. In 
general consideration, the PSF
corresponds to the distributed real response in place of ideally desirable delta function response but in the `same' mathematical 
space. In our case, $f_{ij}$ is in the mathematical space given by density-height distribution  and is generated from a single photon
in spectrum bin, which corresponds to our target space and different from the first one, so we had to 
modify the algorithm as given below.
$$
s_j^{t+1}\sum_i f_{ij}  = s_j^{t}\sum_i\frac{q_i}{c_i}f_{ij},
$$
or,
$$
s_j^{t+1} = s_j^{t}\frac {\sum_i\frac{q_i}{c_i}f_{ij}}{\sum_i f_{ij}}.
\eqno{(8)}
$$
From a uniform set of values, chosen for each of the $s_j^0$s, the algorithm given by Eqs. 8 is 
found to converge, after a few iterations, to a certain distribution of $s_j$s. 
This set of values of $s_j$s can be taken as the injected spectrum.

\subsubsection{Deconvolution using Radial basis function decomposition}

Clearly the basis functions are not regular and orthogonal. 
We can approach this problem in a different way by choosing a set of appropriate radial basis
functions since it can be shown that any continuous function on a 
compact interval can in principle be interpolated with arbitrary accuracy
by a sum of these well behaved functions (Orr, 1996).

Let us divide the height of consideration ($60 - 80$ km) into $N$ equal intervals. We choose $N$ Gaussian functions centered
at the midpoint ($h_n$s) of those $N$ intervals as our radial basis functions. So the $n^{th}$ radial basis function has the form
$$
\phi = e^{\frac {-|h - h_n|}{2\sigma^2}}.
\eqno{(9)}
$$
We divide the whole range of spectrum (say $1-100$ keV) in $M$ separate intervals. We can write each of our original basis as linear combination
of the radial basis functions, so that
$$
f(h,\nu_m) = \sum_n a_{nm}\phi_n,
\eqno{(10)}
$$
where $\nu_m$ corresponds to the $m^{th}$ interval in photon energy. We calculate the coefficients $a_{nm}$ by matrix inversion method
(for example, see Broomhead et al., 1988). Let us assume $a_{nm}$s form a matrix ($A$) of dimension n x m.
As each of the actual basis functions have their major contribution at different altitudes (as they are centered at different heights ) 
the column vectors of matrix $A$ are clearly linearly independent. So the matrix $A$ is invertible.

Now the middle part of Eqs. 4 at time `t' can be written as,
$$
\int I_{0}(\nu, t)f(h, \nu)d\nu = \sum_m I_{0m}\sum_n a_{nm}\phi_n,
$$
$$
= \sum_m \sum_n I_{0m} a_{nm}\phi_n,
$$
$$
= \sum_n (\sum_m a_{nm}I_{0m})\phi_n.
\eqno{(11)}
$$
Similarly, we expand the left hand side of Eqs. 4 at the same time with the radial Gaussian basis functions:
$$
q(h,t) = \sum_n q_n\phi_n.
\eqno{(12)}
$$
Comparing Eqs. 11 and Eqs. 12 we have,
$$
\sum_n a_{nm} I_{0m} = q_n. 
\eqno{(13)}
$$
This is a matrix equation, which can be written as,
$$
AI_0 = Q.
\eqno{(14)}
$$
Inversion of this equation 
$$
I_0 = A^{-1}Q
\eqno{(15)}
$$
gives the spectrum at the time of consideration.

The second method is an one step (matrix inversion) process and the existence of unique deconvolution is highly 
sensitive to the exact evaluation of the basis functions and the values of the parameters, namely $\alpha$ and $\lambda$.
On the other hand, the first method is similar to an iterative spectrum fitting process 
and gives a result irrespective of variation (even large) of the parameter and basis values. 
We have tried both the deconvolution methods and found that the first method gives the
best approximation for the injected spectra of the flares. We have 
considered this method in the current study and working on the optimization of the second one.

\section{Observation and analysis}

In what follows, We analyze VLF amplitudes corresponding to one M2.6 and one X2.2 class solar flares which occurred
on $7^{th}$ of June, 2009 and $15^{th}$ of February, 2011 respectively. We use VLF signal amplitude for the NWC transmitter
received at the Ionospheric and Earthquake research Centre (IERC) of Indian Centre for Space Physics (ICSP) 
at a distance of 5691 km. The phase information has no role in obtaining the injected spectrum.
First, by using the LWPC (Long Wave Propagation Capability) code developed by Ferguson (1998), we obtain
changes in the ionospheric parameter due to solar flares and 
then by using Wait formula (Wait, 1960; Wait, 1962; Wait and Spice, 1964)
we derive electron density profile as a function of time and height 
(Thomson, 1993; Grubor et al., 2008; Pal and Chakrabarti, 2010). The LWPC code is a 
collection of separate programs which can be used according to user
requirement. The Long-Wave Propagation model (LWPM) is used as the default program in
LWPC. This model treats the ionosphere as having exponential increase in conductivity
with height. A log-linear slope ($\beta$ in $km^{-1}$) and a reference height ($h'$)
define this exponential model. We use this default LWPC model to find out the simulated
unperturbed signal amplitude ($A_{lwpc}$) at a particular time.

In Figs. 2(a-b), we plot observed amplitude variations of VLF signal as a function of time during 
the M and X class flares. First, we calculate the deviation of the signal amplitude 
by subtracting the value of quiet day signal amplitude ($A_{quiet}$) 
from the perturbed signal amplitude ($A_{perturb}$) due to flares.
$$
{\Delta}A=A_{perturb}-A_{quiet}.
\eqno{(16)}
$$
We add this $\Delta$A with the unperturbed simulated signal
amplitude as obtained from LWPC.
$$
A^{'}_{perturb} = A_{lwpc}+{\Delta}A.
\eqno{(17)}
$$
This $A'_{perturb}$ is then used to obtain $\beta$ and $h'$
parameters under the flare conditions. For this, we use the
range-exponential model of LWPC where the user provides a set of trial
$\beta$ and $h'$ parameter at each time. The LWPC program is run to
obtain the signal amplitude for that set and this amplitude is compared
with $A^{'}_{perturb}$ (Grubor et al., 2008; Pal et al., 2010). This process
is repeated till LWPC gives the signal amplitude which matches with the
$A'_{perturb}$. The set of $\beta$ and $h'$ which agrees best at a given time is chosen.

The well-known Wait's formula, that has been used to calculate the electron density profile
for lower ionosphere during the flare is given by, 
$$
N_{e}(h)=1.43 \times 10^{13} exp(-0.15h')exp[(\beta-0.15)(h-h')],
\eqno{(18)}
$$
where, $N_{e}$ is the electron density in $m^{-3}$ at a height $h$. We put the deduced
values of $\beta$ and $h'$ to calculate the electron density at different heights (for different $h$ values)
during the peaks of the flares.

In Figure 3, we plot electron densities around the peak time of (a) M-class and (b) X-class flares at
different altitudes in the D-region.
\begin{figure}[h]
\centering{
\includegraphics[height=2.5in,width=2.5in,angle = 270]{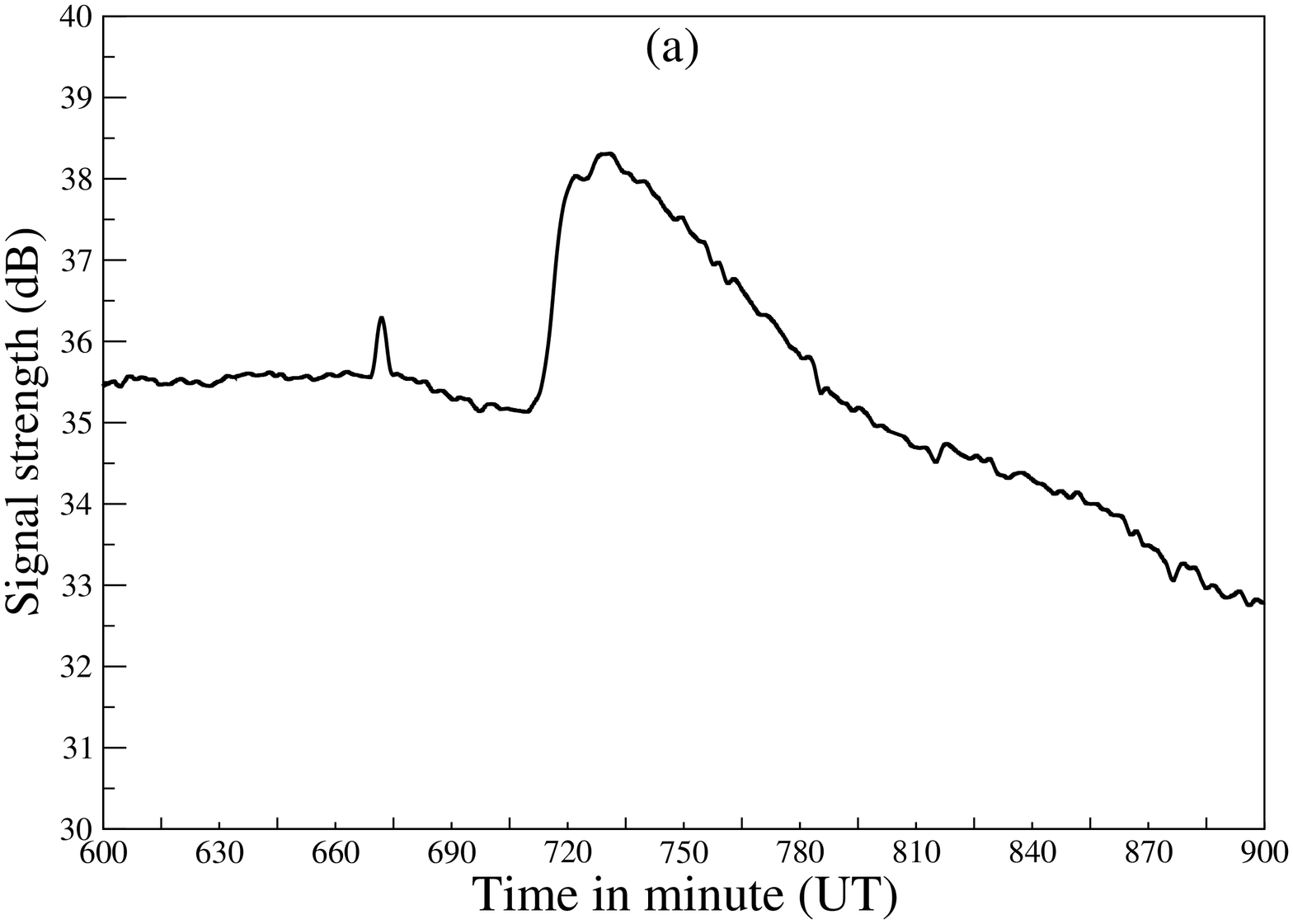}
\includegraphics[height=2.5in,width=2.5in,angle = 270]{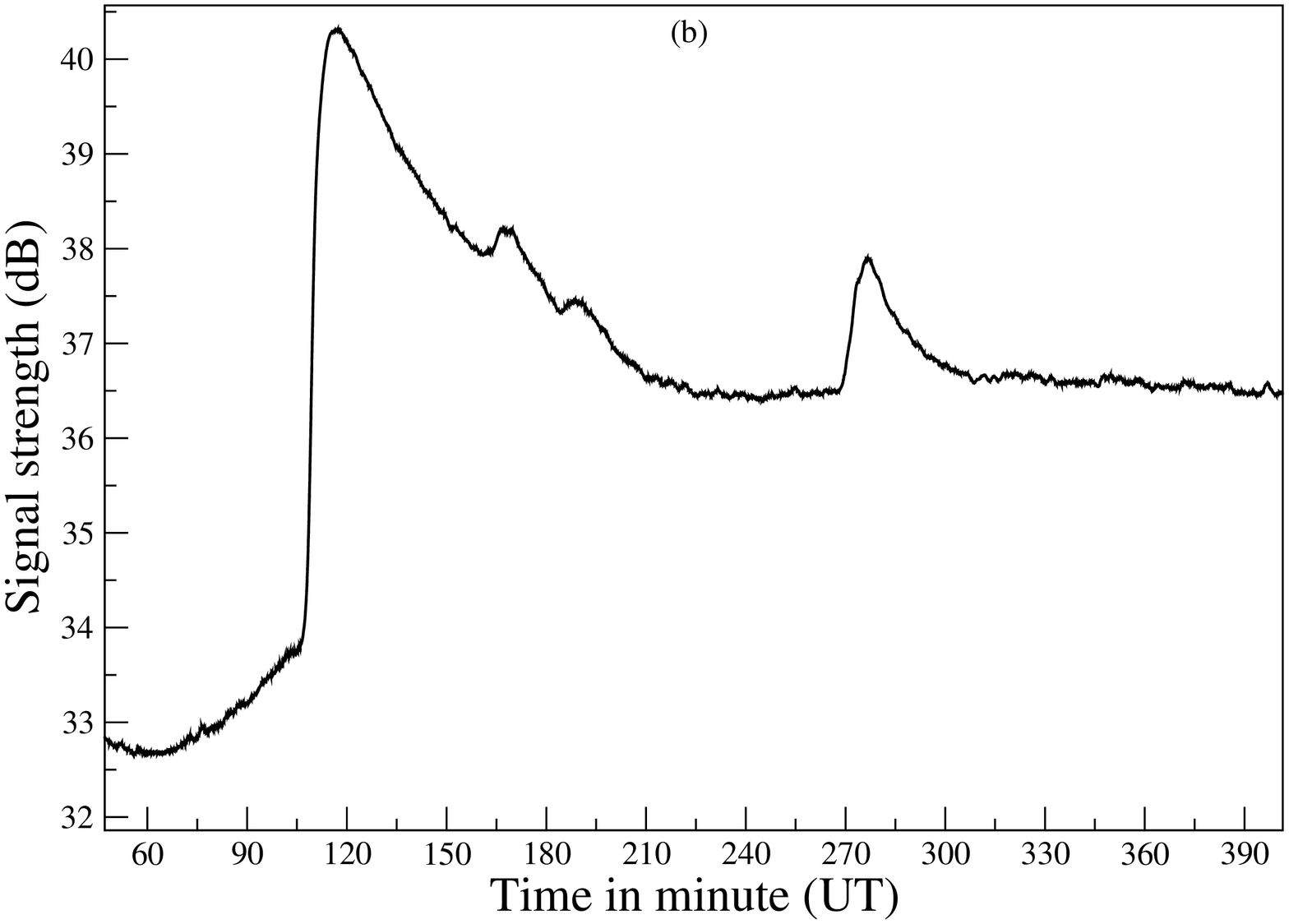}
\caption{Observed VLF amplitude for the (a) M class and the (b) X class flares.}
\label{fig:basis}}
\end{figure}
Values of the function $q(h,t)$, as obtained from Eqs. 4 are shown in Fig. 4.
\begin{figure}[h]
\centering{
\includegraphics[height=2.5in,width=2.1in,angle = 270]{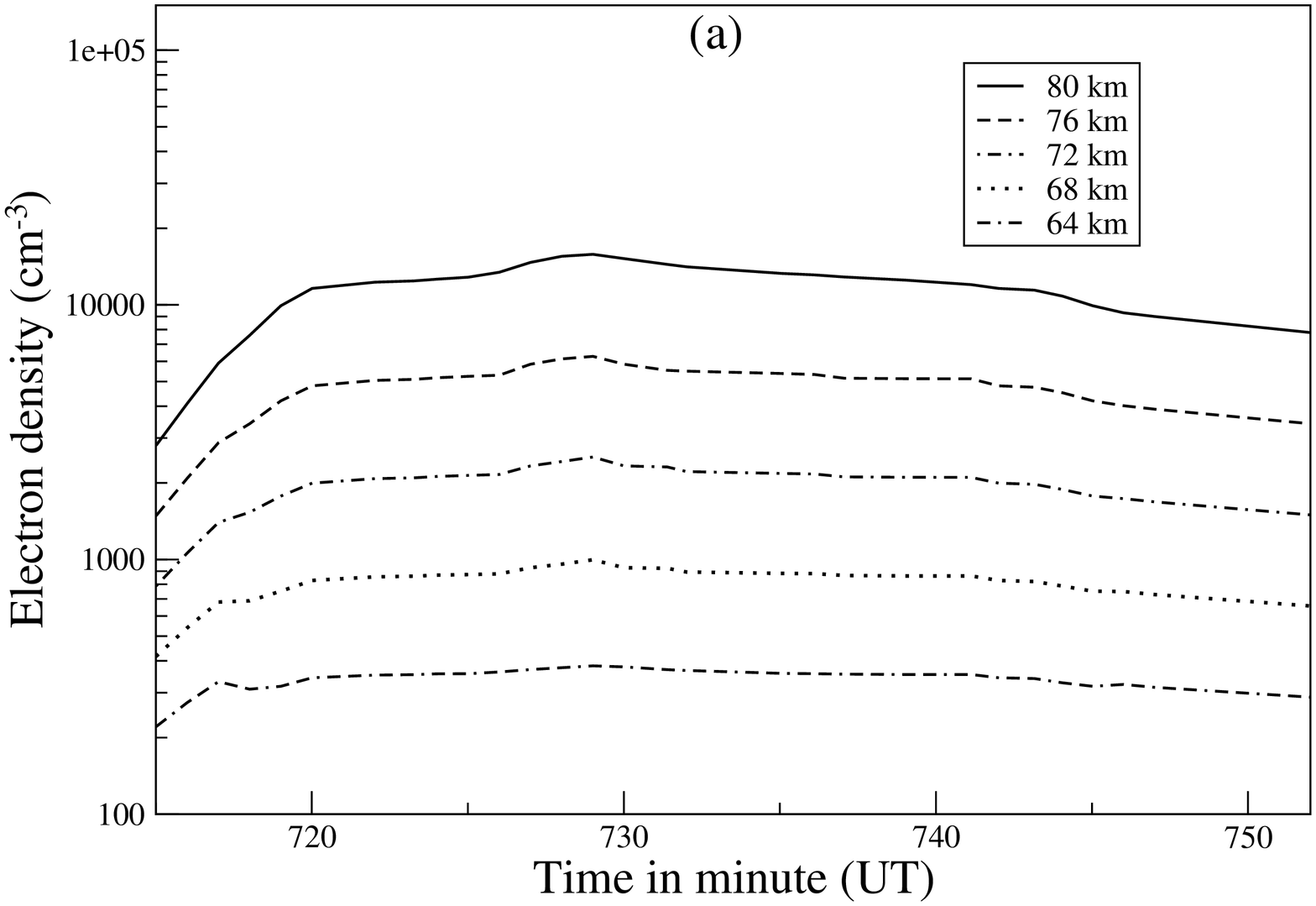}
\includegraphics[height=2.5in,width=2.1in,angle = 270]{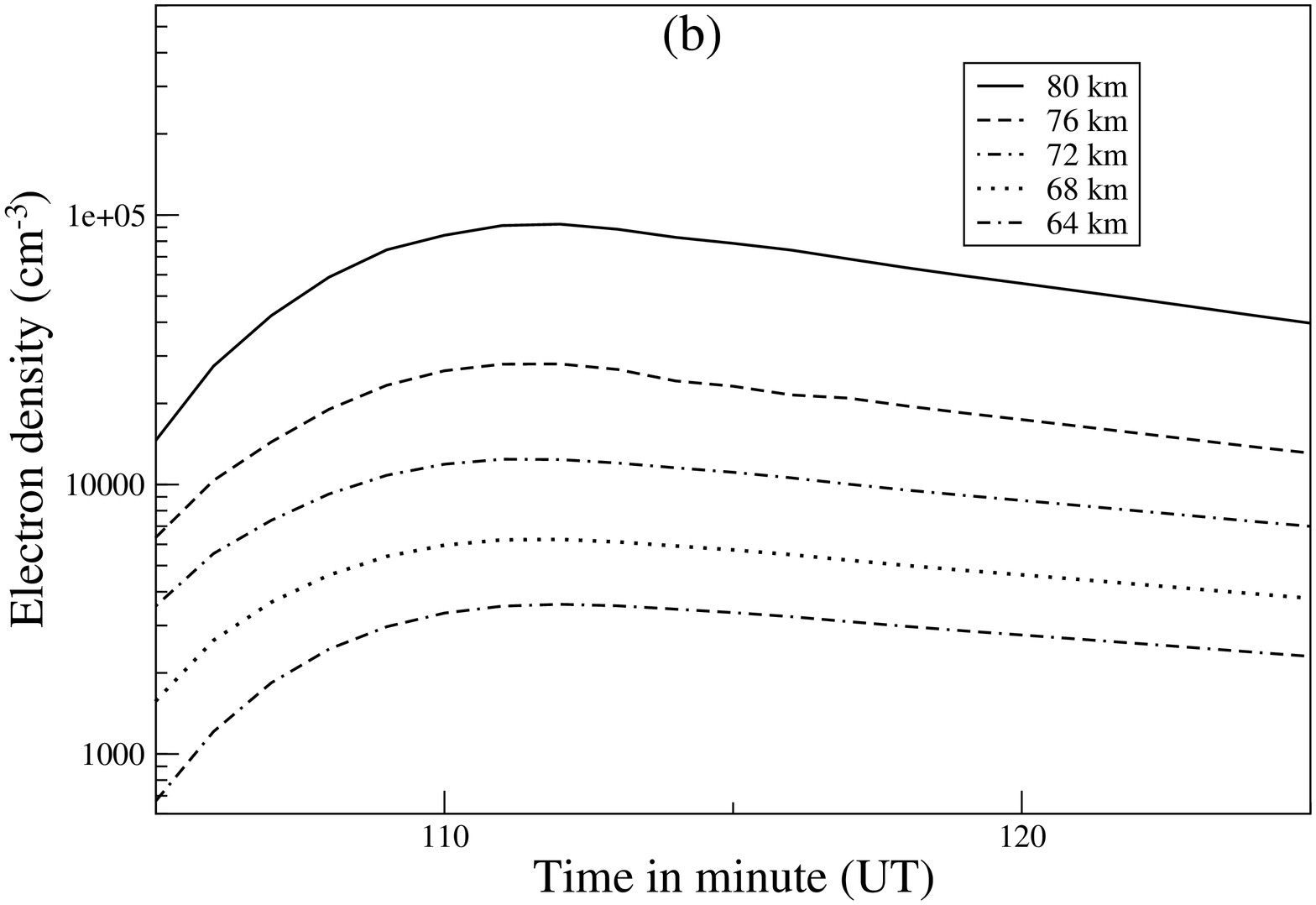}
\caption{Electron densities around the peak of the flares as calculated using 
Wait's formula from the VLF data at different D-region altitude for 
the (a) M-class and (b) X-class flares.}
\label{fig:basis}}
\end{figure}

In our results we observe that peak of the electron density and the peak of the 
ionization rate occur at different times. As discussed in
numerous papers (Zigman et al., 2007; Basak et al., 2013; Palit et al., 2014), the 
electron density peak and hence VLF modulation peak
during any such ionization event should appear after a delay from that of the event peak or the ionization peak. This is due to
the sluggishness of the lower ionosphere. We use the first deconvolution method (Section 2.3.1) 
to calculate the spectra of the two flares at their peaks using the profile of $q(h)$ shown in Fig. 4.

\begin{figure}[h]
\centering{
\includegraphics[height=2.5in,width=2.1in,angle = 270]{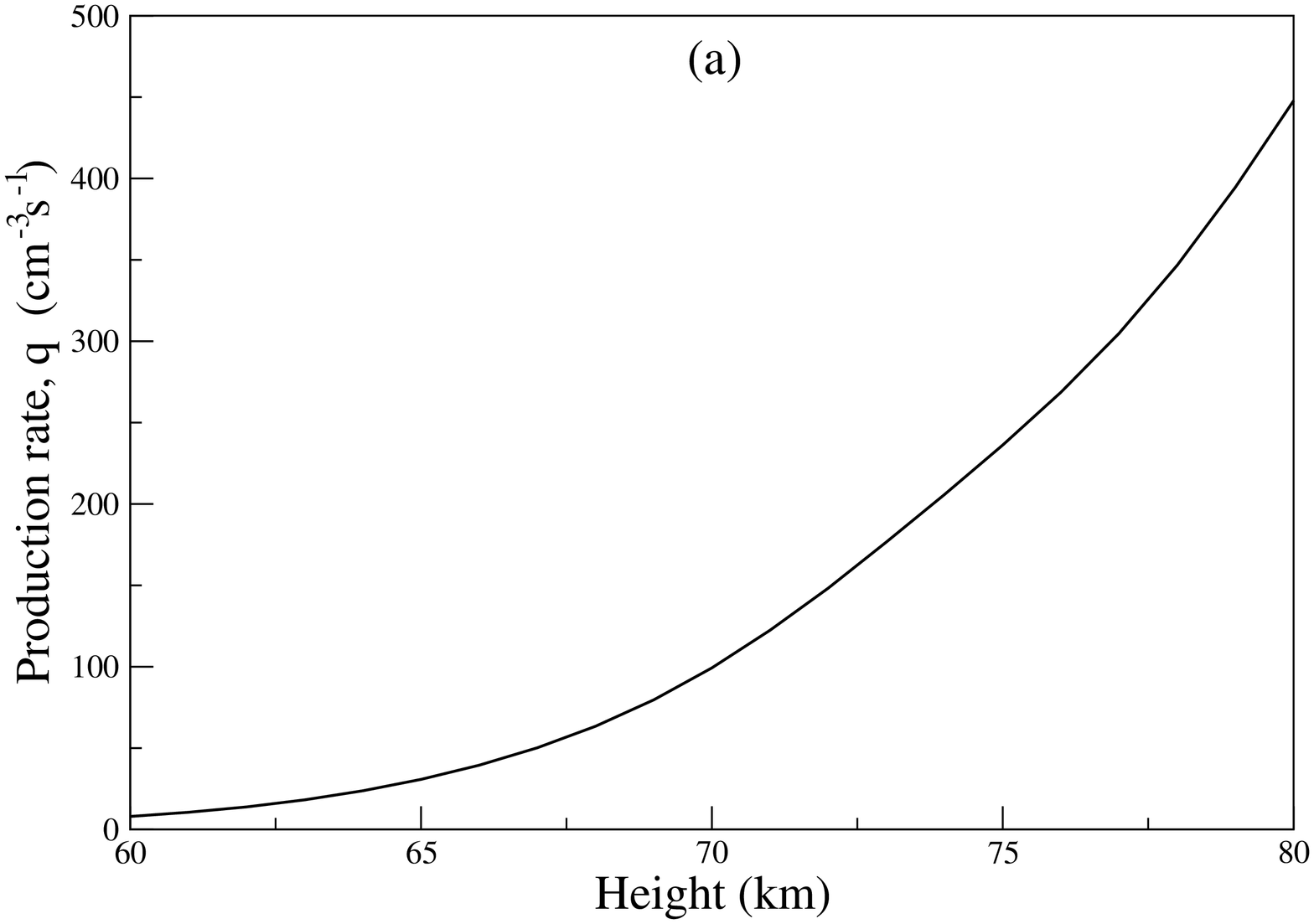} %\hspace{0.1 cm}
\includegraphics[height=2.5in,width=2.1in,angle = 270]{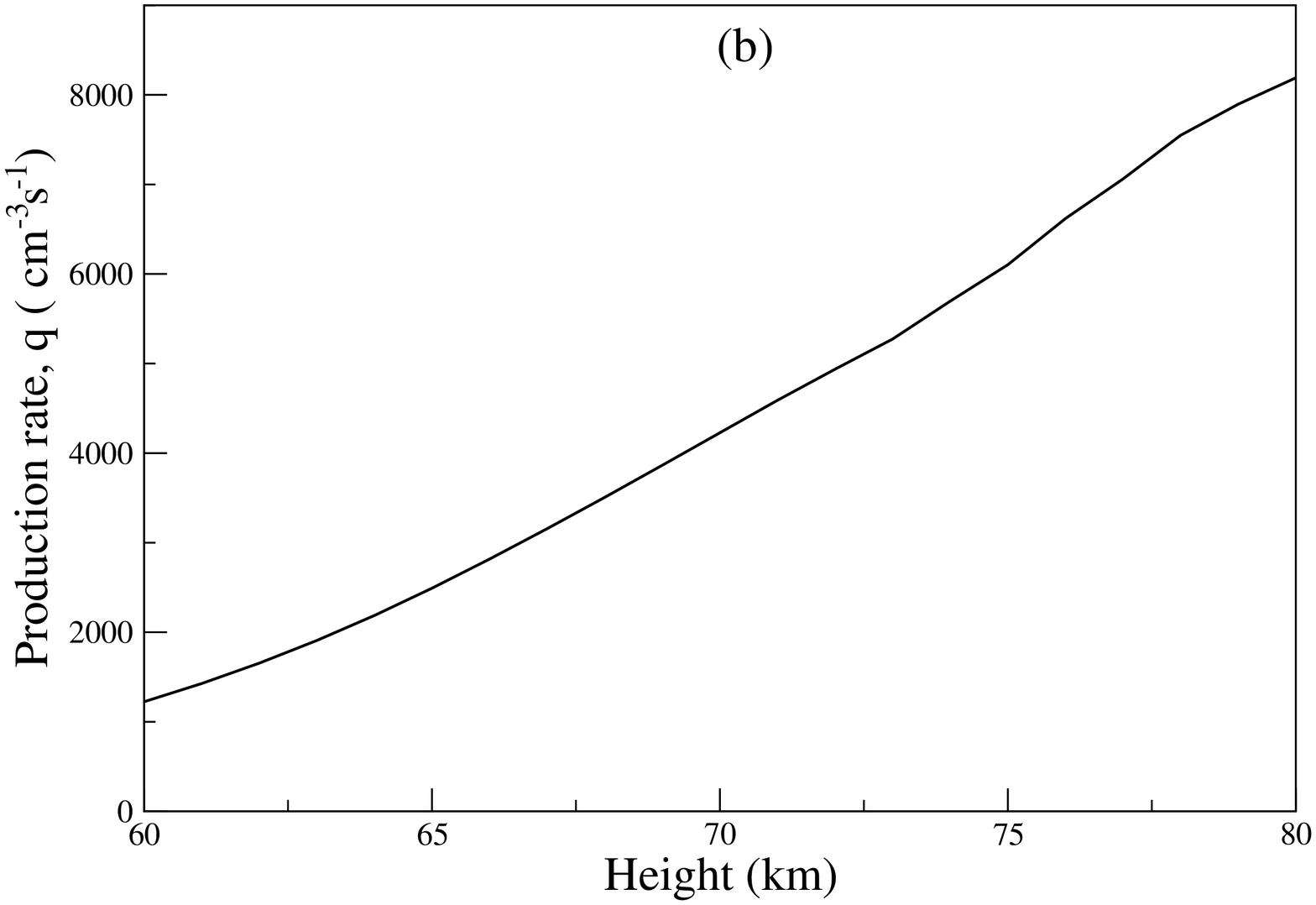}
\caption{Profile of $q(h)$ at peak times of the (a) M-class 
and the (b) X-class flares calculated from electron density profiles
in Fig. 3 using the right hand side of Eqs. 4. }
\label{fig:basis}}
\end{figure}

\section{Results}

In Fig. 5, we show calculated photon spectra at the peak time of the M and the X-class flares under consideration. The solid curves correspond
to the calculated spectra of the flares and the dashed curves correspond to those obtained from RHESSI
satellite data (Sui et al., 2002). We took the spectra
of the flares from RHESSI satellite using {\it sswidl} data analysis software. From the Figure we see that the
reconstructed spectra only  up to $\sim 15$  keV match with the Rhessi spectra satifactorily and above this energy, those deviate 
from the RHESSI spectra significantly. This is probably because in the range of altitude of 60 - 80 km which VLF 
radio wave is sampling, the skywave is affected by soft X-ray photons only. From the nature of the calculated basis functions 
we find that for photons from 4 to 7 keV, most of the ionization occur in the height range of $60-80$ km,
below (2-3 keV) and above (8 - 15 keV) this band, smaller but significant fraction of total ionization occurs 
in that height range. We anticipate that our procedure to obtain an approximate spectrum obtained by
sampling $60-80$ km of the D-layer should be generally acceptable.

Since VLF radio samples only a thin shell of the ionosphere, it produces response function of a
narrow energy band. Thus, it is impossible to reconstruct the whole injected spectrum. 
In Fig. 1 we clearly see that the high energy photons produce maximum number of electrons 
at a much lower height. Our method can reconstruct this high energy component also, only if 
we combine more advanced set of observations, such as those using Incoherent Scatter Radar, 
along with VLF studies. Our present effort is to be viewed as the first attempt to 
use the Earth's lower ionosphere as a gigantic detector to reconstruct major radiation 
component that has impinged on it. In future, we will combine results of multi-wavelength observations 
to reconstruct more complete injected spectrum. Indeed, our method would be very useful for those strong sources
which may even saturate conventional detectors.

\begin{figure}[h]
\centering{
\includegraphics[height=3.3in,width=3.5in,angle = 270]{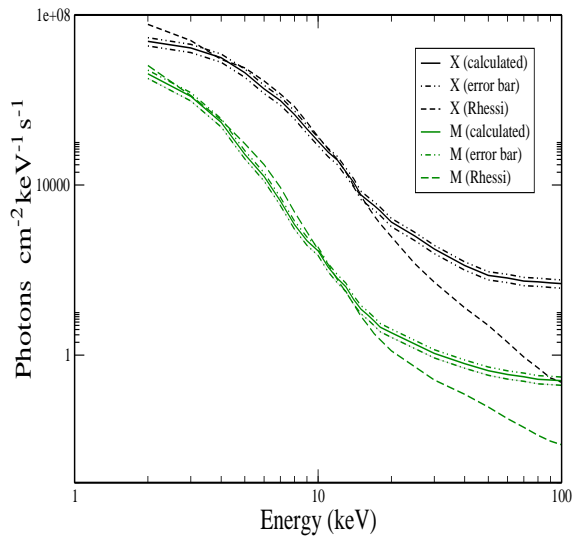}
\caption{Spectra of the flares at the peak times. Solid curves correspond to calculated 
spectra and dashed ones are the corresponding spectra obtained from RHESSI data. Lower pair 
of curves correspond to the spectra of M-class flare and upper pair of curves correspond
to the spectra of X-class flare. Error bars calculated in Section 5 are also plotted with the calculated spetra.}
\label{fig:basis}}
\end{figure}

\section{Observation limit and accuracy of spectra}

In this Section  we estimate the accuracy with which we can estimate a spectrum of a solar flare in the range of our
interest with the VLF observation of ionospheric modulation. The limit on accuracy is imposed by the minimum precision of
the observed VLF data. This calculation is significant in two ways: first, it imposes an error bar on the calculated
spectra and  second, which is more important, it is equivalent to the resolving power of the assumed ionosphere-detector,
(with VLF response analysis as read out mechanism) i.e., the resolution or minimum difference  in spectral features that the
detector can measure.   

Since LWPC follows the Mode
theory of VLF wave propagation in Earth-ionosphere waveguide, we adopt the calculations from this theory.
For the propagation of VLF in the curved path between the Earth's ionosphere and
ground and assuming a uniform height along the path the vertical electric field strength between a vertical transmitter and
receiver is given by (Wait 1960),
$$
E = E_0 W,
\eqno{(19)}
$$
where,
$$
W = (\frac {\frac {d}{a}}{sin (\frac {d}{a})})^\frac {1}{2} (\frac {d}{\lambda})^\frac {1}{2} \frac {\lambda}{h}e^{i(\frac {2\pi d}{\lambda} - \frac {\pi}{4})}\sum_{n=0}^{\infty}\Lambda_n G_n e^{-i2\pi \frac {C_n^2}{2}(\frac{d}{\lambda})}. 
\eqno{(20)}
$$
Here, $E_0$ is the field of the source at a great-circle distance $d$ in a flat perfectly conducting earth, $a$ is the
radius of the Earth, $h$ is the height of the lower edge of the ionosphere and $\lambda$ is the
free-space VLF wavelength. $\Lambda_n$ is the coupling excitation factor between VLF transmitter and different
modes. $G_n$ represents the height gain factor at this wavelength and is normalized to 1
at the ground and $C_n$ is the cosine of angle of incidence of $n^{th}$ wave mode in the lower ionospheric layer.

The square of the amplitude of the field at the receiver can be written as,
$$
|W^2| \sim (\frac {\frac {d}{a}}{sin (\frac {d}{a})}) \frac {d}{\lambda} (\frac {\lambda}{h})^2\sum_{n=1}^{\infty}\sum_{m=1}^{\infty}
e^{-i\frac {k}{2}(C_n^2 - C_m^{*2})d}.
\eqno{(21)}
$$
The VLF signal received, i.e., the value of the power intensity averaged over width of the waveguide is obtained from
$|\overline{E}^2| = |E_0^2||\overline{W}^2|$, where,
$$
|\overline{W}^2| = \frac {1}{h} \int_0^h|W^2|dh
$$
$$
= (\frac {\frac {d}{a}}{sin (\frac {d}{a})}) \frac {d}{\lambda} (\frac {\lambda}{h})^2\sum_{n=1}^{\infty} e^{-i\frac {k}{2}(C_n^2 - C_n^{*2})d}
\eqno{(22)}
$$
Noting
$$
C_n^2 -C_n^{*2} = kIm C_n^2,
\eqno{(23)}
$$
assuming large distance where very few ($\sim$ 1) waveguide mode persist $\alpha d \gg 1)$ and replacing $C_n$ by $C$ we have
$$
|\overline{W}^2| \sim \frac {\frac {d}{a}}{sin (\frac {d}{a})} \frac {d}{\lambda} (\frac {\lambda}{h})^2e^{-\alpha d}
e^{-(2G)^{\frac {1}{2}} \frac {d}{h}},
\eqno{(24)}
$$
where,
$$
\alpha = k Im \frac {C^2}{2} ,
\eqno{(25)}
$$
and
$$
G = \frac {\epsilon\omega}{\delta_g + i\epsilon_g\omega},
\eqno{(26)}
$$
where, the $\epsilon$ and $\epsilon_g$ correspond to the dielectric constant
of the lower ionosphere and ground respectively, $\delta_g$ is the ground
conductivity and $\omega$ is the wave angular frequency (Wait 1964).

If the VLF signal amplitude corresponding to the ambient and flare conditions are $V_a$
and $V_f$ respectively, then the difference
$$
\Delta V = V_f - V_a \approx 20 \times 2.3 \times ln (\frac {\overline{W}_f^2}{\overline{W}_a^2}),
\eqno{(27)}
$$
where, $\overline{W}_a$ and $\overline{W}_f$ are the average power intensity during ambient and flare conditions.

From Eqs. (24) we have,
$$
\Delta V \approx 46.0 \times [2(ln \frac {h_1}{h_2}) + (\frac {1}{h_2} - \frac {1}{h_1}) (2Gd^2)^\frac {1}{2} - (\alpha_2 -\alpha_1)d] 
\eqno{(28)}
$$

The above theory is for an assumed sharply bounded ionosphere, the effect of exponential variation of ionospheric
conductivity can be included through the modification of the attenuation rate parameter $\alpha$.
For VLF wave mode propagation the amplitude part of the mode resonance equation reads as
$$
e^{(-2d\alpha)} = |R_i||R_g|
\eqno{(29)}
$$
where $R_i$ is the reflection coefficient of the lowest layer of the ionosphere and $R_g$ is that of the ground.

If $R_{i2}$ and $R_{i1}$ are the ionospheric reflection coefficients during the disturbed and ambient conditions respectively and corresponding attenuation coefficients are $\alpha_2$ and $\alpha_1$ respectively, then
$$
\alpha_2 - \alpha_1 = -\frac {1}{2d} ln|\frac {|R_{i2}|}{|R_{i1}|}.
\eqno{(30)}
$$
For lower D region the effective dielectric constant of the medium can be approximated by an exponential function, such that the relative
permittivity can be put in the form (Wait, 1963),
$$
K(h) = K_0(1-i\frac{1}{L}e^{\beta h}),
\eqno{(31)}
$$
where, $K_0$ is the reference permittivity and $L$ is a constant.
The parameter $\beta$ is known as the conductivity gradient of the ionosphere and is given by,
$$
\beta = 2.3 \frac {log (\frac {\sigma}{\sigma_0})} {(h - h_0)},
\eqno{(32)}
$$
where, $\sigma$ and $\sigma_0$ are the conductivity at height $h$ and a reference height $h_0$ respectively.
Then it can be shown that the amplitude of the reflection coefficient for $n^{th}$ mode for any type of polarization can
roughly be expressed in the form,
$$
|R_i|= e^{(-\frac {2\pi^2}{\lambda_0 \beta}C)}.
\eqno{(33)}
$$
From Eqs. (30) and (33) we can see that,
$$
\alpha_2 - \alpha_1 = \frac {\pi^2}{\lambda_0 d C}(\frac {1}{\beta_2} - \frac {1}{\beta_1}).
\eqno{(34)}
$$
Putting in Eqs. 28 we get,
$$
\Delta V \approx 46 \times  [2(ln \frac {h_1}{h_2}) + (\frac {1}{h_2} - \frac {1}{h_1}) (2Gd^2)^\frac {1}{2} - \frac {\pi^2}{\lambda_0  C}(\frac {1}{\beta_2} - \frac {1}{\beta_1}). 
\eqno{(35)}
$$

In the process of finding the parameters $h^\prime$ and $\beta$ from VLF observation (Section, 3)
during disturbed condition uncertainty may appear in the values of the parameters.
So replacing $\beta_2$ by $\beta$ and $h_2$ by $h^\prime$ and taking the differential we get,
$$
2\delta (\Delta V) \approx 46 \times  [(-\frac {2}{h^\prime} - \frac {(2Gd^2) ^\frac {1}{2}}{h^{\prime2}}) \delta h^\prime + \frac {\pi^2}{\lambda_0  C} \frac {1}{\beta^2}\delta \beta] .
\eqno{(36)}
$$
The factor of 2 in the left hand side of the equation appears due to the fact that error or uncertainty may appear during the
observational measurements of the VLF signal for both ambient and flare situations. In our case the minimum
precision, i.e., the maximum error with which we can observe the VLF signal is $\delta (\Delta V) =0.1 dB$.
The maximum uncertainty in $h^\prime$ and $\beta$ i.e., $\delta h^\prime$ and
$\delta \beta$ can be calculated by putting the remaining terms equal to zero for each case.

Now taking logarithm and then differential of Eqs. 18 we get
$$
\frac {\delta N_e}{N_e} = (h-h^\prime)d\beta + (0.65 - \beta)\delta h^\prime
$$
or
$$
\delta N_e = \gamma N_e .
\eqno{(37)}
$$

Putting the maximum uncertainties in the values of calculated parameters obtained from Eqs. 36 in Eqs. 37 we can find the maximum
uncertainty in the calculated electron density as a function of height.

Now taking differential of Eqs. 4 and substituting the value of $\delta N_e$ from Eqs. 37 we have
$$
\delta q = \int \delta I_{0}(\nu, t)f(h, \nu)d\nu =  \gamma (1+ \lambda)(\frac {dN_e}{dt} + 2\alpha N_e^2)
\eqno{(38)}
$$
The quantity $\delta I_{0}(\nu, t)$ can be calculated with the iterative  maximum likelihood method as described in Section 2.3.1.
Two error bars can be put to the calculated spectra by adding and subtracting the values of $\delta I_{0}(\nu, t)$ with the
calculated  $I_{0}(\nu, t)$ values.
With the value of $d = 5690 km$, calculated value of $G = 4.6 \times 10^{-4}$, $\lambda_0 = 15 km$, $C \sim 0.1$
and corresponding values of
$h^\prime$ and $\beta$ for the flares we find the values of $\gamma$ for the M and the X-class flares are  respectively $1.1\times 10^{-2}$ and $.9\times 10^{-2}$. The error bars calculated with these values from Eqs. 38 are plotted in
Fig. 5 with the corresponding calculated spectra.

\section{Conclusions}

In this paper, we have demonstrated that the Earth's ionosphere may be treated as a detector for strong 
sources of high energy radiation. Depending on the physics of the interaction of the 
probe with the ionosphere, the energy range of the spectrum  will vary.
In the present situation we use VLF radio propagation in the Earth-ionosphere waveguide which samples lower ionosphere
(60-80km) during moderately strong flares. Since this layer is most sensitive to photons of up to about $12$ keV, the spectra 
obtained at higher energies are not accurate enough. We find (Fig. 5) that the Thermal bremsstrahlung emission in M and X-class
flares have been reproduced up to about $15$keV quite satisfactorily. It is to be noted that in Palit et al (2013), we have
already solved the `forward problem', i.e., reproduction of the deviation of VLF signals obtained by VLF antennas from 
the X-ray spectrum. To our knowledge, the `inverse problem', namely, predicting radiation spectrum from 
radio sonde studies is carried out for the first time.  We are aware of the fact that our final result strongly depends on 
the electron number density derived from the signal amplitude anomaly. The conventional and widely used LWPC
code assumes that this number density is uniform over the entire propagation path and thus depending on  
a specific path, where such assumption is clearly violated, other method has to be adopted to compute 
electron density measurement. Our method uses all the tools available, namely, 
(a) $h'$ and $\beta$ parameters obtained at the peak flare time using LWPC, (b) the electron number density N and dN/dT
using Wait's formula at times close to the peak and finally (c) height distribution of the ion production rate from 
continuity equation and its derivatives. Thus the accuracy of the final result can be improved only after improving
these intermediate steps.

Presently satellites engaged in observing solar radiation have severe limitations. GOES obtains light curves 
in two wide bands in X-rays. RHESSI misses spectra at peaks of many flares as its good-time of observation is about twenty minutes 
per hour. In that respect a series of relatively less expensive ground based antennas which can observe the Sun round the clock 
would be more than welcome, since accurate behavior of the solar spectrum can be obtained treating the Earth as a gigantic
detector. Ours is thus a novel and alternative approach to obtain spectral information from strong sources with a 
maintenance free large area detector.

The accuracy of the process adopted here, however, depends on the following important points.

\noindent  1. Choice of accurate basis function set is crucial. The basis functions  
are calculated here from purely theoretical considerations. For a proper
deconvolution, the basis should be adjusted from the knowledge of the ionization obtained from theoretical and observational
study. Simulations such as Monte Carlo method adapted to calculate the ionization in Palit et al. (2013) are
required. In time, the procedure could be perfected.

\noindent 2. Proper calculation or observation of electron density distribution over height and time should be done. The LWPC
estimation of the electron density is assumed on the empirical Wait model of lower ionosphere. Accuracy of the  
deconvolution of spectrum depends on the accurate estimation of electron density distribution at all heights.
For this one may have to combine VLF observations with other methods, such as RADAR measurements. In any case,
our approach remains valid and result would be more accurate when methods to obtain electron density is perfected.

\noindent 3. Knowledge of the chemical processes and prevalent rate coefficients and their evolution 
throughout the flare occurrence time would be required. For instance, for weaker flares new reactions requiring lower photon
energy are to be fed in computing the basis functions.

Once we have probes to sample different layers of ionosphere, we are in a position to obtain
a complete and more accurate description of the injected spectrum of all combined 
events on the ionosphere. In the day time, the dominating ionizing agent is the Sun itself and therefore
it is expected that solar spectrum would be more accurately obtained, than the spectra of distant energetic events.
One can find solar flare spectra in the entire energy range by extending the method above D-layer
for the UV band of the spectrum and below D-layer for the hard X-ray and gamma ray bands of the spectrum. 
Spectra of more energetic events such as gamma ray bursts and soft gamma ray repeaters  in real time may 
be obtained, even when a satellite is saturated or misses the peak data. We shall explore these possibilities in future.

%%\end{linenumbers}

\acknowledgments
Sourav Palit acknowledges MoES for financial support.

%% To help institutions obtain information on the effectiveness of their
%% telescopes, the AAS Journals has created a group of keywords for telescope
%% facilities. A common set of keywords will make these types of searches
%% significantly easier and more accurate. In addition, they will also be
%% useful in linking papers together which utilize the same telescopes
%% within the framework of the National Virtual Observatory.
%% See the  Web site at http://aastex.aas.org/
%% for information on obtaining the facility keywords.

%% After the acknowledgments section, use the following syntax and the
%% \facility{} macro to list the keywords of facilities used in the research
%% for the paper.  Each keyword will be checked against the master list during
%% copy editing.  Individual instruments or configurations can be provided 
%% in parentheses, after the keyword, but they will not be verified.

%%{\it Facilities:} \facility{Nickel}, \facility{HST (STIS)}, \facility{CXO (ASIS)}.

%% Appendix material should be preceded with a single \appendix command.
%% There should be a \section command for each appendix. Mark appendix
%% subsections with the same markup you use in the main body of the paper.

%% Each Appendix (indicated with \section) will be lettered A, B, C, etc.
%% The equation counter will reset when it encounters the \appendix
%% command and will number appendix equations (A1), (A2), etc.

%%\appendix

%%\section{Appendix material}

\clearpage


\begin{thebibliography}{}
\bibitem{Afraimovich2001} Afraimovich E. L., Altynsev A. T., Grechnev V. V., Leonovich L. A., Ionospheric effects
of the solar flares as deduced from global GPS network data, Adv. Space Res. Vol. 27, Nos 6-7, pp. 1333-1338, 2001.

\bibitem {Basak2013} Basak, T., Chakrabarti, S. K., Effective recombination coefficient and solar zenith angle effects
on low-latitude D-region ionosphere evaluated from VLF signal amplitude and its time delay during X-ray solar flares
Astrophys Space Sci. 2013, doi 10.1007/s10509-013-1597-9



\bibitem {Broomhead1988} Broomhead D. S., Lowe D., Radial Basis Functions, Multi-Variable Functional Interpolation and
Adaptive Nctworks. Royal Signals and Radar Establishment Memorandum 4148, March 28, 1988.




\bibitem {Chakrabarti2011} Chakrabarti S. K., Mondaln S. K., Sasmal S., Bhowmick D., Chowdhuri A. K., Patra N. N., First VLF
detections of ionospheric disturbances due to Soft Gamma ray Repeater SGR J1550-5418 and Gamma Ray Burst GRB 090424,
Indian J. Physics, 84(11) 1461-1466, 2010



\bibitem {Chapman1931} Chapman, S., Absorption and dissociative or ionising effects of monochromatic radiation in an atmosphere
on a rotating earth, Proc. Phys. Soc., London, 43, 1047-1055, 1931


\bibitem{Ferguson1998} Ferguson, J.A. Computer programs for assessment of long-wavelength radio communications, version 2.0.
Technical document 3030, Space and Naval Warfare Systems Center, San Diego, 1998

\bibitem{Glukhov1992} Glukhov V.S., Pasko V.P., Inan U.S., Relaxation of transient lower ionospheric disturbances caused by
lightning-whistler-induced electron precipitation bursts; Journal of Geophysical Research, 97:16971, 1992





\bibitem{Grubor2005} Grubor D. P., Suliic D. M., and Zigmann V., Influence of solar flares on the
Earth ionosphere waveguide, Serb Astron. J. No. 171, 29-35, 2005, doi:10.2298/SAJ0571029G




\bibitem{Grubor2008} Grubor D. P., Suliic D. M.,  and Zigmann V., Classification of X-ray solar flares regarding their effects
on the lower ionosphere electron density profile, Ann. Geophys., 26, 1731-1740, 2008.





\bibitem{Hedin1991} Hedin A. E., Extension of the MSIS Thermosphere Model into the middle and lower atmosphere;
Geophysical Research Letters 96, NO. A2, P. 1159, 1991.



\bibitem{Hubbell1995} Hubbell J. H. and Seltzer S. M., Tables of X-RAY Mass Attenuation Coefficients and Mass 
Energy-Absorption Coefficients 1 keV to 20 MeV UCRL-50400, Vol. 6, Rev. 5 26 EPDL97UCRL-50400, Vol. 6, Rev. 5
EPDL97 for Elements Z = 1 to 92 and 48 Additional Substances of Dosimetric Interest," NISTIR 5632, National 
Institute of Standards and Technology, 1995.



\bibitem{Inan2007} Inan U. S., Lehtinen N. G., Moore R. C., Hurley K., Boggs S., Smith D. M., Fishman G. J.; Massive
disturbance of the daytime lower ionosphere by the giant g-ray flare from magnetar SGR 1806-20 ; Geophysical
Research Letters, vol.34, L08103, 2007

\bibitem{Liu2004} Liu J. Y., Lin C. H., Tsai H. F., Liou Y. A., Ionospheric solar flare effects monitored by the
ground-based GPS receivers: Theory and observation, Journal of Geophysical Research, 2004, doi: 10.1029/2003JA009931.




\bibitem{Lucy1974} Lucy R. B., An iterative technique for the rectification of observed distributions. The Astronomical
Journal, 79(6):745–754, 1974


\bibitem{McRae2003} Mcrae W. M. and Thomson N. R., Solar flare induced ionospheric d-region enhancements from VLF phase
and amplitude observations, Journal of Atmospheric and Solar Terrestrial Physics, 66(1), 77-87,2003, doi:10.1016/j.jastp.2003.09.009.




\bibitem{Mitra1972} Mitra A. P. and Rowe J. N., Ionospheric effects of solar flares-VI. Changes in D-region ion chemistry during
solar flares, Journal of Atmospheric and terrestrial Physics, Vol-34, pp 795-806. Permagon Press, Northrrn ireland, 1972




\bibitem{Mondal2012} Mondal K. S., Chakrabarti S. K., Sasmal S., Detection of ionospheric perturbation due to a soft gamma ray
repeater SGR J1550-5418 by very low frequency radio waves; Astrophys Space Sci, 341:259–264, 2012, doi 10.1007/s10509-012-1131-5


\bibitem{Orr1996} Orr M. J. L.; Introduction to Radial Basis Function Network, Centre for Cognitive Science, Edinburgh University, EH
9LW, Scothland, UK, 1996 (http://www.anc.ed.ac.uk/ \& mjo/rbf.html).


\bibitem {Palit2013} Palit S., Basak T., Pal S., Chakrabarti S. K., Modeling of very low frequency (VLF) radio wave signal profile due
to solar flares using the GEANT4 Monte Carlo simulation coupled with ionospheric chemistry, Atmos. Chem. Phys., 13, 9159-9168, 
2013, doi:10.5194/acp-13-9159-2013


\bibitem{Palit2014}  Palit S., Basak T. Pal S., and  Chakrabarti S. K., Theoretical study of lower ionospheric response to solar 
flares: Sluggishness of D­region and Peak time delay;  Astrophys Space Sci 355:2190, 2014i, doi 10.1007/s10509­014­2190­6


\bibitem {Pal2010} Pal S. and Chakrabarti S. K., "Theoretical models for Computing VLF wave amplitude and phase and their 
applications", AIP, 1286, 42, 2010



\bibitem {Rees1989} Rees M. H., Physics and Chemistry of the upper atmosphere, Cambridge Univ Press, Cambridge, Great Britain, 1989





\bibitem{Richardson1972}  Richardson W. H., Bayesian-based iterative method of image restoration. Journal of the Optical Society
of America, 62(1):55–59, 1972



\bibitem {Sojka2013} Sojka J. J., Jensen J., David m., Schunk R. W., Woods T., Eparvier F., Modeling the ionospheric E and F1 regions: Using SDO-EVE
observations as the solar irradiance driver, Journal of Geophysical Research: SPACE PHYSICS, VOL. 118, 5379–5391, doi:10.1002/jgra.50480, 2013




\bibitem {Solomon2006} Solomon C. S., Numerical models of the E-region ionosphere, Advances in Space Research 37, 1031–1037, 2006

\bibitem {Sui2002} Sui  L., Holman G.D. et al.,“Modeling Images and Spectra of a Solar Flare Observed by RHESSI on 20
February 2002”, Kluwer Academic Publishers, 2002


\bibitem {Tanaka2008} Tanaka Y.T., Terasawa T., Yoshida M., Horie T., Hayakawa M.; Ionospheric Disturbances caused
by SGR 1900+14 giant gamma ray flare in 1998: Constraints on the energy spectrum of the flare, GR, VOL.113, A07307, 2008




\bibitem {Tanaka2010} Tanaka Y. T.,  Raulin J. P., Bertoni F. C. P., Fagundes P. R., Chau J.,  Schuch N. J., Hayakawa M., Hobara Y,
Terasawa, T., and Takahashi T., First Very Low Frequency detection of short repeated bursts from Magneter SGR J1550–5418,
ApJ -721 L24, 2010, doi:10.1088/2041-8205/721/1/L24




\bibitem {Taylor1970} Taylor G. N. and Watkiks C. D., Ionospheric Electron Concentration Enhancement during a Solar Flare,
Nature 228, 653 - 654 (14 November 1970); doi:10.1038/228653a0.


\bibitem {Thomson1993} Thomson N. R., "Experimental daytime VLF ionospheric parameters", J. Atmos. Sol.-Terr. Phys., 55(2) , 173, 1993

\bibitem {Turunen1992} Turunen E., Matveinen H., Ranta H., Sodankyla Ion Ccemistry(SIC) model, Sodankyla Geophysical Observatory report No.
49, Sodankyla, Finland, 1992

\bibitem{Wait1960} Wait J. R., Terrestrial propagation of very-low-frequency radio waves: A theoretical investigation, 
J. Res. Natl. Bur. Stand., Sect. D, 64(2), 153–203, 1960

\bibitem{Wait1962} Wait J. R., Excitation of modes at very low frequency in the Earth-ionosphere wave guide, J. Geophys. Res., 67(10), 1962

\bibitem{wait1963} Wait J. R., Walter L. C., Reflection of VLF Radio Waves From an Inhomogeneous Ionosphere. Part 1. Exponentially Varying 
Isotropic Model, 1963, Journal of research of national Bureau of standerds-D. Radio propagation, Vol 67D. No. 3. May-June-1963

\bibitem{wait64} Wait, J. R. \& Spies, K. P., Characteristics of the earth-ionosphere waveguide for
VLF radio waves, NBS Tech. Note 300, 1964

\bibitem {Watanabe2013} Watanabe D., Nishitani N., Study of ionospheric disturbances during solar flare events using the SuperDARN
Hokkaido radar, Advances in Polar Science,  Vol. 24 No. 1: 12-18, March 2013, doi: 10.3724/SP.J.1085.2013.00012.

\bibitem{Whitten1965} Whitten R. C. \& Poppoff I. G., Physics of the Lower Ionosphere, Prentice- Hall, Inc. Engle-wood Cliffs, N. J., 1965

\bibitem {Xiong2005} Xiong B., Weixing W., Liu L., Withers P., Zhao B., Ning B., Wei Y., Le H. J., Ren Z., Chen Y., He M., Liu J.,
Ionospheric response to the X‐class solar flare on 7 September 2005; JGR, vol. 116, A11317, doi:10.1029/2011JA016961, 2011

\bibitem {Zigman2007} Zigman, V., Grubor, D., Sulic, D.,  D-region electron density evaluated
from VLF amplitude time delay during X-ray solar flares. J. Atmos. Terr. Phys., 69, 775-792. 2007





\end{thebibliography}
\end{document}